\documentclass[%
 aip,
 jcp,
 a4paper,
 12pt,
 amsmath,amssymb,
preprint,%
]{revtex4-1}

\usepackage{graphicx}
\usepackage{dcolumn}
\usepackage{bm}
\usepackage{amssymb}
\usepackage{graphicx}
\usepackage{subfig}
\usepackage{epstopdf}
\usepackage{color} 
\usepackage{ae}
\usepackage{setspace}
\usepackage{makeidx}
\usepackage{float}
\usepackage[none]{hyphenat}
\usepackage{mathtools}

\usepackage[utf8]{inputenc}
\usepackage[T1]{fontenc}
\usepackage{mathptmx}

\begin{document}

\preprint{AIP/Journal of Chemical Physics}

\title{Water diffusion in rough carbon nanotubes}

\author{Bruno H. S. Mendon\c{c}a}
\email{brunnohennrique13@gmail.com}
\affiliation{Instituto de F{\'i}sica, Universidade Federal do Rio Grande do Sul, Porto Alegre, 
RS 91501-970, Brazil}

\author{Patricia Ternes}
\affiliation{Campus S{\~a}o Bento do Sul, Instituto Federal Catarinense, S{\~a}o Bento do Sul, SC, 
89283-064, Brazil}

\author{Evy Salcedo}
\affiliation{Coordenadoria Especial de F{\'i}sica, Qu{\'i}mica e Matem{\'a}tica, Universidade Federal de Santa Catarina, Ararangu{\'a}, SC, 88905-120, Brazil}

\author{Alan B. de Oliveira}
\affiliation{Departamento de F{\'i}sica, Universidade Federal de Ouro Preto, Ouro Preto, MG, 35400-000, Brazil}

\author{Marcia C. Barbosa}
\affiliation{Instituto de F{\'i}sica, Universidade Federal do Rio Grande do Sul, Porto Alegre, 
RS 91501-970, Brazil}

\date{\today}

\begin{abstract}
We use molecular dynamics simulations to study the diffusion of water inside deformed carbon nanotubes with different degrees of 
deformation at $300$ K. We found that the number of hydrogen bonds that water forms depend on nanotube topology, leading to 
enhancement or suppression of water diffusion. The simulation results reveal that more realistic nanotubes should be considered 
to understand the confined water diffusion behavior, at least for the narrowest nanotubes, when the interaction between water 
molecules and carbon atoms are relevant.
\end{abstract}

\maketitle

\section{Introduction}

Physical and chemical properties of materials change dramatically depending on the confining size and geometry~\cite{RAT03,FON19}. 
For large systems, in the thermodynamic limit, surface effects are irrelevant when computing bulk properties. This is not the case 
of nanoscale confining geometries where volumetric and surface interactions have a similar order of magnitude. Then the nature of 
the surface becomes  a relevant factor for computing  thermodynamic and dynamic properties of the confined fluid. One example of a 
surface impact on the fluid phase is water confined in carbon nanotubes (CNTs). Studies focusing fluid transport in carbon nanotubes 
are ubiquitous~\cite{HUM01,MAJ05,WHI07}. 
Even though water transport properties in perfect nanotubes already 
present a number of interesting phenomena, 
deformations in nanopores change the degree of 
 confinement which impacts the transport properties. These changes are present 
 both in  artificial and biological systems. For instance, deformations are relevant for the study of DNA sequencing~\cite{ELG18}, 
 carbon nanotubes as nanosyringes~\cite{HOL04} and nanothermometers~\cite{LIS02}. 

Water exhibits a number of anomalous properties already in the bulk. When confined in nanostructures, additional anomalies 
arise~\cite{ALE08,HAR80,YIN08,FAR11,YON12,KOH19,KOH19X}. Water confined in carbon nanotubes exhibits a non-monotonic behavior 
diffusion coefficient. For diameters above 6 nm the diffusion coefficient approaches to the bulk water value~\cite{FAR11}. For 
diameters smaller than 6 nm, the diffusion coefficient increases slightly as the CNT diameter decreases, reaching a maximum 
value for CNT diameters close to 2.5 nm~\cite{FAR11}. For even smaller diameters, the diffusion coefficient decreases with 
decreasing diameter, reaching a minimum value for diameters close to 1.2 nm, followed by a significant increase in diffusion 
in the narrowest nanotubes. This non-monotonic behavior was observed in simulations~\cite{JUN04,KOL04,NAG04} and this high 
mobility observed in experiments~\cite{MAM06}.

The diffusion coefficient of nanoconfined water depends on  two competing interactions: hydrogen bonds between water molecules and 
the water-wall interaction. While the water hydrogen bonds decreases the energy by forming a tetrahedral network, which decreases 
the mobility, the water-wall interactions are less attractive molecules, creating  dangling bonds, which privilege their mobility. 
Therefore, the precise nature of the water-wall interaction can affect the diffusion. For instance, theoretical and experimental 
works have also shown that different tube chiralities affect the confined water distribution inside carbon nanotubes and 
consequently its diffusion~\cite{CHA06,WAN04,ANS13}. 

Some studies indicate that armchair nanotubes have larger diffusivity  when compared with the zigzag~\cite{YIN05,YIN08}. These 
studies, however, were made of very small tubes and the diffusion was computed for short times what might have affected their 
results. The changes in the  length of the tubes produce a large impact on the mobility~\cite{JUN04,ALE08}. This suggests that 
comparison between different chiralities should be performed for larger tubes.

Since the water diffusion inside carbon nanotubes depends strongly on the nanotube diameter~\cite{BAO11,GAU04}, the defects which 
decrease of the nanotube radius might impact the mobility of the water. In the production of the nanotubes, these defects and 
contaminations are expected~\cite{OLI16}. The resulting distortions from the perfect nanotube structure might lead to changes 
in the thermodynamic and dynamic behavior of the confined fluid. 

Recently, a theoretical study analyzed the impact of a uniformed deformation in the diffusion coefficient of confined water. 
This study employed the TIP4P/2005 water model confined in a  kneaded armchair nanotube~\cite{BHSM18}. This specific type of 
deformation leads to the reduction in the lateral space which enhanced the number of hydrogen bonds and consequent reduced the 
water diffusion coefficient for almost all $(n,n)$ nanotubes. The exception was $(9,9)$ CNT where water has a near 
zero diffusion when the nanotube has zero deformation.
In this particular case, deformation promotes the disruption of the hydrogen bond network. 
The ice-like structures formed for the (9,9) perfect nanotube melts by the compression, improving water mobility. Even though 
interesting, the study of water inside kneaded nanotubes is not realistic since it preserves the axial symmetry while nanotubes 
with randomly distributed deformations, wrinkled nanotubes, do not exhibit any specific symmetry~\cite{IIJ91,IIJ93}. 

In this paper, we analyze through molecular dynamics simulations the effects of nonuniform deformations and chirality in the 
mobility of confined water inside carbon nanotubes. We compare the diffusion coefficient for the TIP4P/2005 water model when 
the liquid is confined in an armchair and zigzag wrinkled nanotubes. Our results for the two types of nanotubes are analyzed 
for perfect, kneaded and wrinkled tubes. 
The remaining of this paper goes as follows. In Section II the model is presented. Results are discussed in Section III, while 
Conclusions are shown in section IV.

\section{Model and Methods}

\begin{figure}[H]
  \begin{center}
  \includegraphics[width=6.4in]{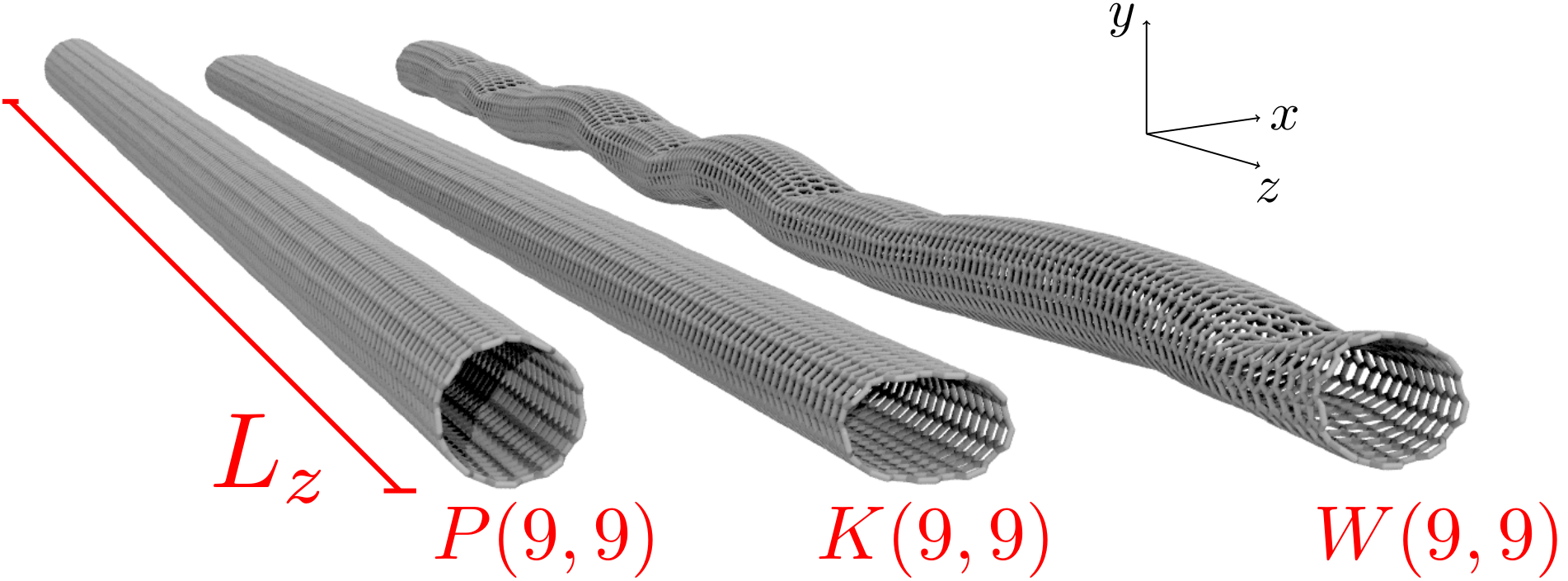}
  \end{center}
  \caption{Snapshot of perfect, P$(9,9)$, kneaded, K$(9,9)$ and wrinkled, W$(9,9)$. nanotubes}
  \label{fig_modelo}
\end{figure}

We studied the structural and dynamic behavior of  water confined in carbon nanotubes with two different types of deformation: 
kneaded and wrinkled. In order to produce the kneaded nanotubes, a perfect nanotube was uniformly kneaded in the $y$ direction 
until the nanotube reached an elliptical cylinder shape with eccentricity of $0.8$, as illustrated in the 
Figure~\ref{fig_modelo}~\cite{BHSM18}. The deformation imposed to the  kneaded nanotube is characterized by the eccentricity:
\begin{equation}
\label{eq:dist0029}
e= \sqrt{1 - \frac{b^{2}}{a^{2}}}
\end{equation}
\noindent where $b$ is the smaller semi-axis and $a$ is the largest semi-axis. 

The second process comprises to disorderly compress the nanotube in the $y$ direction until it forms nonuniform wrinkles. 
Wrinkles in the systems were created in a disorderly manner, but as the number of wrinkles was small, the size distribution of 
the segments between two wrinkles was the same for nanotubes of the same diameter. In average, each nanotube was compressed in 
$5$ diferent $z$ positions, with  with values of $e$ ranging from $0.0$ to $0.8$. 

In addition, for comparison purposes, perfectly structured nanotubes were also produced. The perfectly symmetrical nanotubes 
are characterized by a $0.0$ eccentricity. Figure~\ref{fig_modelo}  represents  the three types of nanotubes used: perfect, 
kneaded and wrinkled.

In order to illustrate the difference in water behavior due to structural changes in the ($n,m$) nanotube, two different diameters 
of armchair ($n=m$) and zigzag ($m=0$) carbon nanotubes were used. Details of the structure of each nanotube are given in 
Table~\ref{tab_CNT}. Note that $(9,9)$ and $(16,0)$ CNTs have approximately the same diameter. This is also the case for the 
$(12,12)$ and $(21,0)$ nanotubes.

\begin{table}[H]
\begin{center}
\caption{\label{tab_CNT}Parameters for perfect carbon nanotubes.}
\begin{tabular}{ccccc}
\hline \hline
CNT $\quad$ & $\quad$ d ($nm$) & $\quad$ L$_{z}$ ($nm$) & $\quad$ H$_{2}$O & $\quad$ $\rho$ ($g/cm^{3}$) 
\\ \hline
(9,9) $\quad$  & $\quad$ 1.22  & $\quad$ 50.5  & $\quad$ 908   & $\quad$  0.92 \\    
(12,12)$\quad$ & $\quad$ 1.63  & $\quad$ 22.5  & $\quad$ 901   & $\quad$ 0.94 \\ 
(16,0)$\quad$  & $\quad$ 1.25  & $\quad$ 50.5  & $\quad$ 908   & $\quad$  0.80 \\
(21,0) $\quad$ & $\quad$ 1.64  & $\quad$ 22.9  & $\quad$ 901   & $\quad$ 0.86 \\
\hline \hline
\end{tabular}
\end{center}
\end{table}

Molecular dynamics (MD) simulations at a constant number of particles $N$, volume $V$ and  temperature $T$ were employed to 
investigate the diffusion of TIP4P/2005 water model~\cite{ABA05}. Periodic boundary conditions in the tube axial ($z$) direction 
and a cutoff radius of $12$ \AA\ were used. We represented the non-bonded interactions (carbon-oxygen) by the Lennard-Jones 
potential with parameters $\epsilon_{CO}=0.11831$ kcal/mol and $\sigma_{CO}=3.28$~\AA~\cite{HUM01}. Interaction between carbon 
and hydrogen was set to zero.  

Water density inside carbon nanotubes was determined considering the excluded volume due to Lennard-Jones interaction between 
carbon in the nanotubes and oxygen atoms in the water. Therefore the effective density will be $\rho= 4M/[\pi(d-\sigma_{CO})^2L_z]$, 
where $M$ is the total water mass into the tubes, while $L_z$ is the nanotube length~\cite{BOR12}.

Simulations were carried out with LAMMPS package~\cite{LAMMPS}. We employed the Particle-Particle Particle-Mesh method to compute 
the long-range Coulomb interactions. This method treats long range interactions and the Coulombian field of real charges in a 
way that can interfere with your own images~\cite{OST17}. We work around this issue by creating an x-y simulation box around 
$100$~nm for all nanotubes, preventing the carbon nanotube from interacting with its own images and avoiding overlapping 
virtual images with real images, minimizing possible errors in method application. The structure of water molecules was 
constrained through the SHAKE algorithm with a tolerance of 1$\times10^{-4}$. The water temperature was maintained at $300$~K 
through the Nos{\'e}-Hoover thermostat with a damping time of $100$~fs and $1$~fs time interval. In all simulations, we keep 
the nanotubes rigid and keep them from getting off the plane. This procedure has been employed in several similar simulations 
which have shown that considering the nanotube as a rigid system is a very reasonable approximation when compared to the case 
where the thermostat is applied throughout the system~\cite{FAR11,HAN06,KOT04}. The system was equilibrated during $5$ ns, 
then properties were stored every $0.01$ ns during $5$ ns, giving a total simulation time of $10$ ns. 

Due to system dimensions, the diffusion is minimal in the radial direction and only the axial diffusion is considered. It is 
determined through the one dimensional Einstein relation as

\begin{eqnarray}
\label{eq:MSD}
D_z = \lim_{\tau \to \infty} \frac{1}{2}\frac{d}{d\tau}\left\langle z^2(\tau) 
\right\rangle,
\end{eqnarray}

\noindent where $\left\langle z^2(\tau)\right.\rangle=\left\langle\left[z(\tau_0-\tau)-z(\tau_0)\right]^2\right\rangle$ is the 
mean square displacement (MSD) in the axial direction. 

In order to characterize the structure of water, we calculated the number of hydrogen bonds (HB) and also made colormaps for 
oxygen occurrence in the $ xy $, $ xz $, and $ yz $ planes. Hydrogen bonds were computed if both of the following geometrical 
criteria were satisfied:\cite{JOS08}

\begin{eqnarray}
\nonumber
 && \alpha\leq 30^{\circ} \\ \nonumber
 && |\vec{r}_{OO}|\leq 3.50 \mbox{ \AA},
 \label{eq_hb}
\end{eqnarray}

\noindent where $\alpha$ is the $OH\cdots O$ angle and $|\vec{r}_{OO}|$ is the distance between two oxygens. 

Oxygen occurrence colormaps were obtained dividing the correspondent plane in square bins of length $0.1$ \AA\ and counting 
the number of oxygens in each square. Higher oxygen densities are represented in red while low densities tend to darker blue tones. 

\section{Results and discussion}

\begin{figure}[H]
\begin{center}
\includegraphics[width=6.4in]{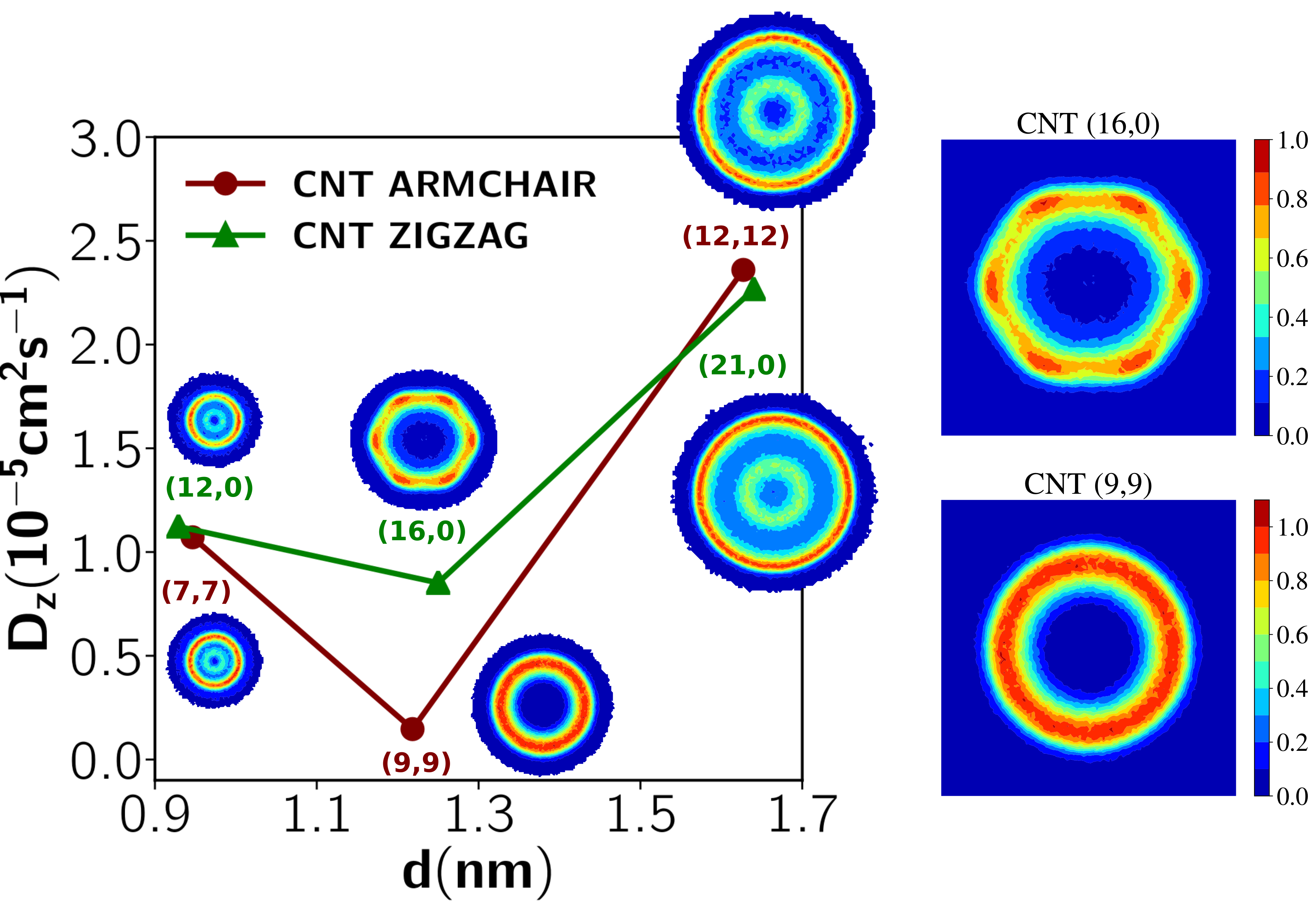}
\end{center}
\caption{Left panel: Diffusion coefficient versus nanotube diameter for perfect P$(n,m)$ armchair and zigzag nanotubes. 
Right panel: $xy$ colormaps show the density of water inside the nanotubes.}
\label{fig_diff}
\end{figure}

\begin{figure}[H]
\begin{center}
{\includegraphics[width=5.4in]{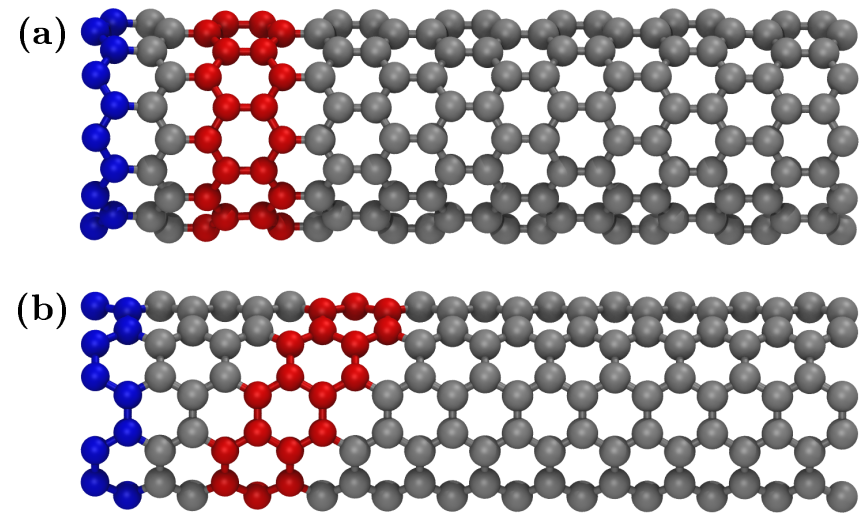}}
\end{center}
\caption{Snapshot of carbon nanotube (a) zigzag and (b) armchair. The structure highlighted in red refer to the zigzag carbon 
rings and armchair spiral-like structure.}
\label{fig_tubosaz}
\end{figure}

First, we compare the mobility of water inside perfect armchair and zigzag nanotubes for different diameters. Figure~\ref{fig_diff} 
shows that the water mobility is not strongly affected by the chirality of the nanotube for different diameters  with the exception 
of the  $(9,9)$ and  $(16,0)$ nanotubes. This diameter is quite distinct when compared with smaller and larger nanotubes because just one water layer close to 
the wall is formed. This water layer behaves quite different depending on the chirality. 
While the water diffusion in armchair $(9,9)$ nanotube is almost null, nonzero diffusion is observed in the zigzag case. The density colormap illustrated in the Figure~\ref{fig_diff} shows that for the $(9,9)$ nanotube water molecules are uniformly 
distributed in the vicinity of the wall. For the $(16,0)$ the fluid molecules  assume a hexagonal distribution at the wall boundary. 
The difference in the water distribution in the two types of nanotubes  is due to the differences in the wall structures as shown 
in the Figure~\ref{fig_tubosaz} combined with the hydrophobic nature of the carbon-water interaction. The water molecules form 
hydrogen bonds but avoid being close to the carbon molecules. 

For the zigzag nanotube one hydrogen bond can be formed between two water molecules located at the middle of two neighbor carbon 
rings at the same $z$ coordinate, forming a ring. For the armchair the hydrogen bond links water in the middle of  two neighbor 
carbon rings, but located at different $z$ coordinates, forming a spiral. The armchair spiral-like structure produces a  more 
connected hydrogen bond network when compared  with the ringlike structure of  the zigzag nanotubes resulting in a lower mobility 
of the armchair $(9,9)$ nanotube.

Next, we address the question of how nanotube deformations affect the water diffusion coefficient. In particular, we analyze the 
difference in water diffusion for armchair and zigzag tubes when they change from perfect to wrinkled and to kneaded. We analyze 
two nanotube diameters for each chirality: $(9,9)$ and $(12,12)$ for the armchair and $(16,0)$ and $(21,0)$ for the zigzag 
nanotubes. The diameter selection was made to test how the compression affects the number of layers and the distinct mobility 
between the two chiralities.

\begin{figure}[H]
\begin{center}
\includegraphics[width=6.4in]{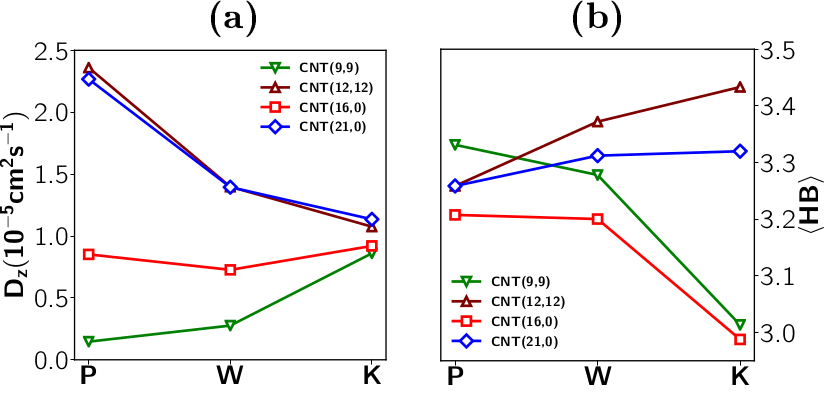}
\end{center}
\caption{(a) Axial diffusion constant and (b) average number of hydrogen bond per water molecule for the nanotubes 
$(9,9)$, $(12,12)$, $(16,0)$ and $(21,0)$. For the cases Perfect P$(n,m)$, Wrinkled W$(n,m)$ and Kneaded K$(n,m)$.}
\label{fig_difHB}
\end{figure}

Deformation changes the way water molecules connect and move within the carbon nanotube, facilitating or hindering the formation 
of hydrogen bonds. This effect directly affects the diffusion of confined water molecules in the nanotube. If we analyze this 
effect locally, we can observe that curvature decreases the number of hydrogen bonds, which can be seen in Figure~\ref{fig_difHB}. 
The Figure~\ref{fig_difHB} shows that for the $(12,12)$ armchair nanotube case deformation decreases axial diffusion by increasing 
the number of hydrogen bonds as confirmed by previous findings~\cite{FAR11,BHSM18}. The same behavior is observed in the $(21,0)$ 
zigzag nanotube. Nonuniform deformations, as in wrinkled nanotubes, bring water molecules  closer to each other, favoring the 
formation of hydrogen bonds as one can see in Figure~\ref{fig_difHB}(b). This effect is even more prominent in kneaded nanotubes, 
in which deformation and decrease of distance between  molecules is more uniform. Changes in  water diffusion due to deformations 
in $(12,12)$ and $(21,0)$ tubes are very similar and chirality seems to play  a minor role in this case.

For armchair $(9,9)$ nanotube, a different scenario is found. Water mobility increases with deformation, which breaks 
hydrogen bonding networks leading to increased confined water diffusion, as can be seen in Figure~\ref{fig_difHB}. For the 
$(16,0)$ zigzag 
nanotube, the  water diffusion and the number of  hydrogen bonds are not affected by the change from perfect to wrinkled nanotube. 
When the nanotube is kneaded, however, the number of bonds is reduced and a clear decoupling between mobility and hydrogen 
bond network disruption is observed. The behavior of the diffusion coefficient and HB of water confined at the $(9,9)$ and $(16,0)$, 
perfect and  wrinkled nanotubes are quite distinct what indicates that for this small diameter  surface effects are indeed relevant. 
However, for the kneaded nanotube water diffusion and the number of hydrogen bonds for both $(9,9)$ and $(16,0)$ nanotubes are 
almost the same indicating that the strong deformation shows stronger impact in the mobility than chirality.

\begin{figure}[H]
  \begin{center}
  \includegraphics[width=6.4in]{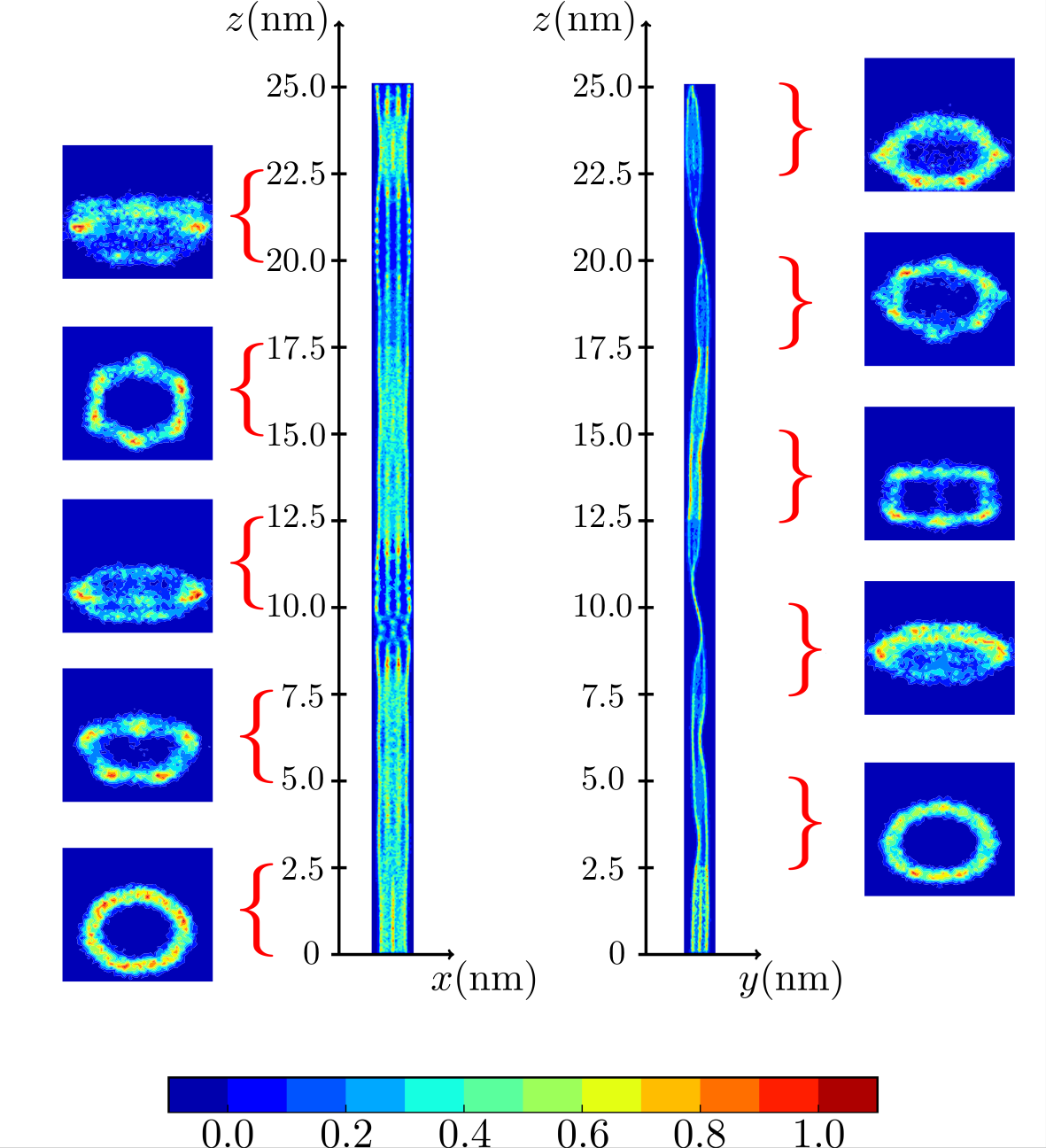} 
  \caption{Colormaps of $xz$, $yz$ and $xy$ directions for the $(9,9)$ wrinkled nanotube. 
           The $xy$ colormaps are dependent on the nanotube region, so each $xy$ 
           colormap is related to one nanotube region with $25$ nm in $z$ direction. Dark blue regions have low probability to 
           find water molecules, while red regions have high probability to find water molecules.}
  \label{fig_wrinkled9}
  \end{center}
\end{figure}

\begin{figure}[H]
  \begin{center}
  \includegraphics[width=6.4in]{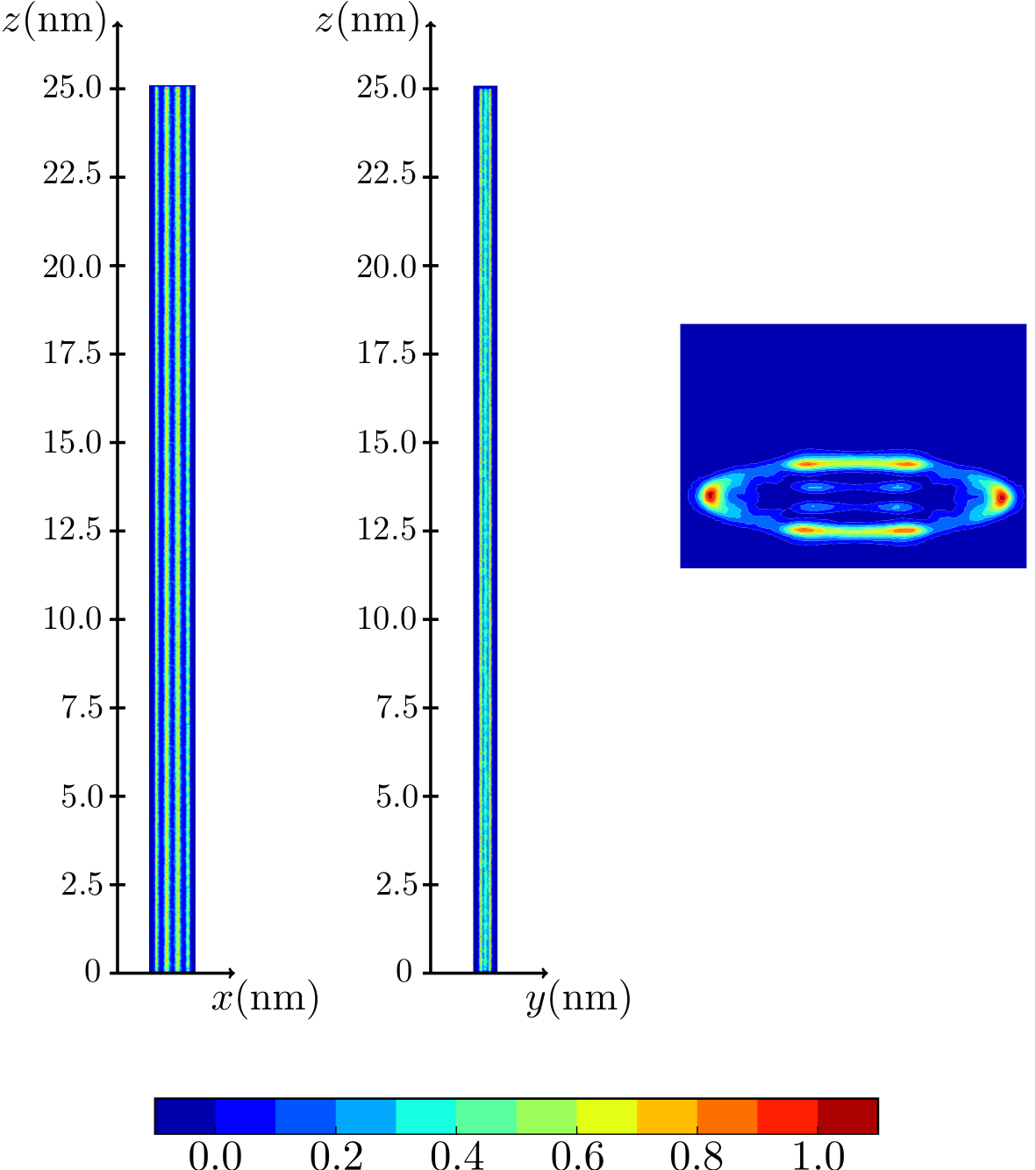} 
  \caption{Colormaps of $xz$, $yz$ and $xy$ directions for the $(9,9)$ kneaded nanotube, $e=0.8$. 
           The $xy$ colormaps are dependent on the nanotube region, so each  $xy$ 
           colormap is related to one nanotube region with $25$ nm in $z$ direction. Dark blue regions have low probability to 
           find water molecules, while red regions have high probability to find water molecules.}
  \label{fig_kneaded9}
  \end{center}
\end{figure}

In order to understand the structural origin of the difference in mobility and HB between wrinkled $(9,9)$ and $(16,0)$ nanotubes, 
we analyzed $xy$, $xz$ and $yz$ colormaps. Figure~\ref{fig_wrinkled9} illustrates these three color maps for the water inside the 
wrinkled $(9,9)$ nanotube. For $25$ nm length nanotube presents perfect segments with no deformation followed by deformed structures 
ranging form different values of $e$ as indicated by Eq.~\ref{eq:dist0029}. Water in a perfect segment is evenly distributed, as 
observed at $0<z<25$ nm in Figure~\ref{fig_kneaded9}, what is consistent with spiral-like behavior with molecules bonded along 
the spiral. As the nanotube becomes deformed these bondings are broken and the molecules in these segments form bonds in the $xy$ 
and $xz$ directions, forming structures quite similar to the perfect $(16,0)$ nanotube. For very strong deformation, the $xy$ 
bonding are destroyed and the molecules form bonded lines in the $xz$ plane.

\begin{figure}[H]
  \begin{center}
  \includegraphics[width=6.4in]{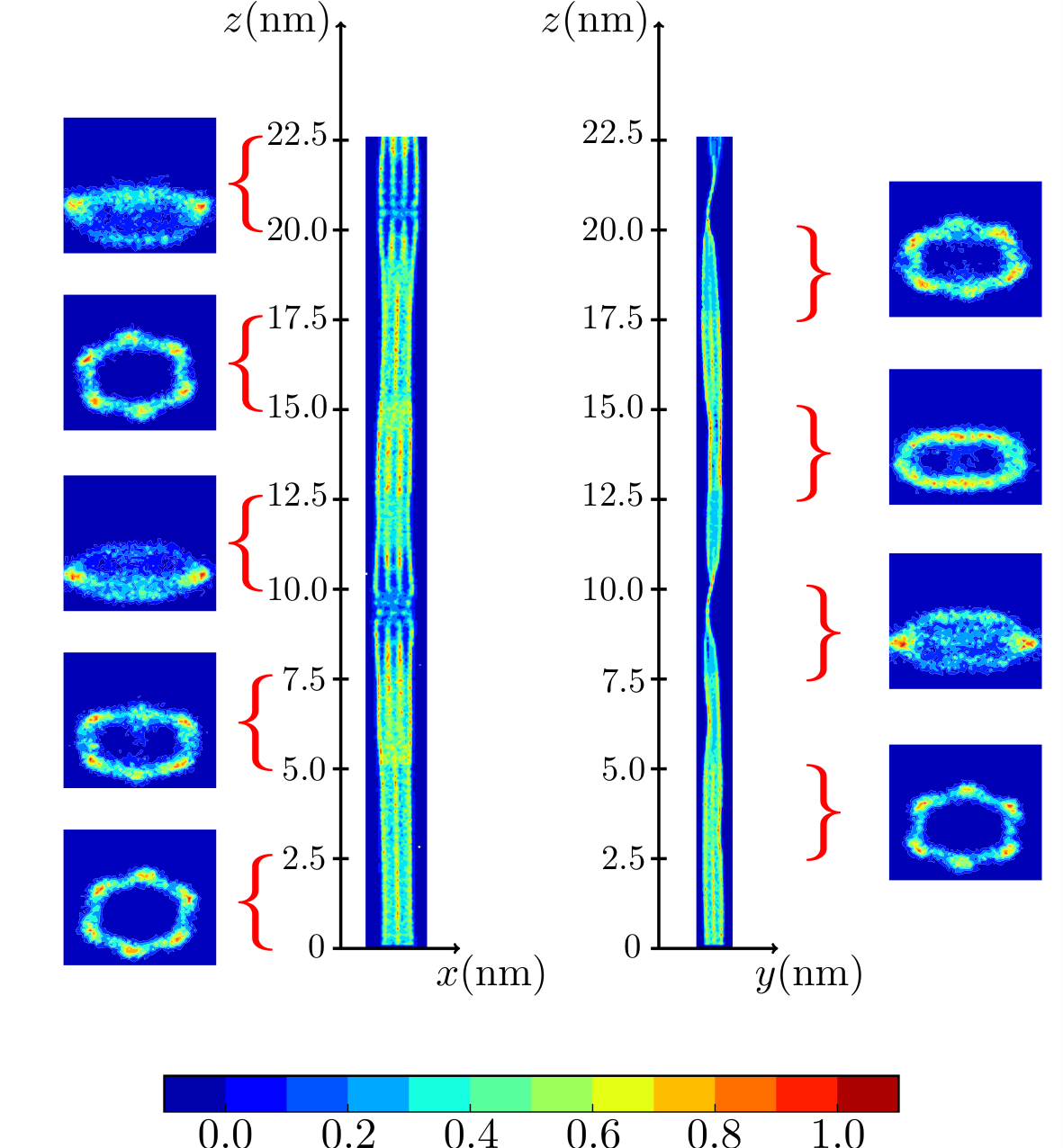} 
  \caption{Colormaps of $xz$, $yz$ and $xy$ directions for the $(16,0)$ wrinkled nanotube. 
           The $xy$ colormaps are dependent on the nanotube region, so each  $xy$ 
           colormap is related to one nanotube region with $22.5$ nm in $z$ direction. Dark blue regions have low probability to 
           find water molecules, while red regions have high probability to find water molecules.}
  \label{fig_wrinkled16}
  \end{center}
\end{figure}

For the $(16,0)$ wrinkled nanotube shown in the Figure~\ref{fig_wrinkled16}, the perfect and not strongly deformed segments 
show a very similar structure, what explains why the random deformation has a very little impact in the mobility and number of the 
hydrogen bonds. The particles show the same structure of bonds and circular mobility observed in the perfect zigzag nanotube. 

\begin{figure}[H]
  \begin{center}
  \includegraphics[width=6.4in]{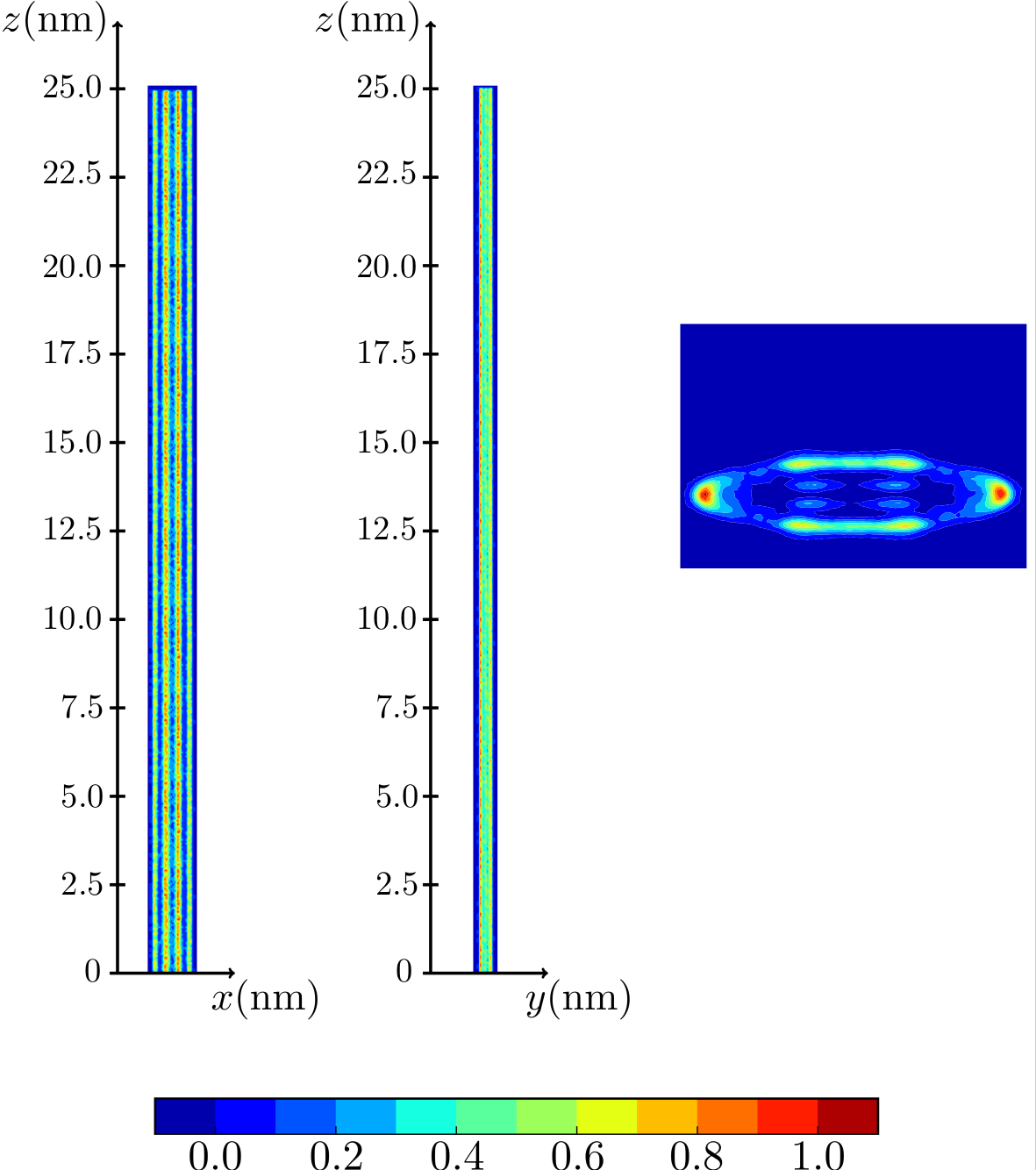} 
  \caption{Colormaps of $xz$, $yz$ and $xy$ directions for the $(16,0)$ kneaded nanotube, $e=0.8$. 
           The $xy$ colormaps are dependent on the nanotube region, so each  $xy$ 
           colormap is related to one nanotube region with $25$ nm in $z$ direction. Dark blue regions have low probability to 
           find water molecules, while red regions have high probability to find water molecules.}
  \label{fig_kneaded16}
  \end{center}
\end{figure}

For the kneaded armchair and zigzag deformed nanotubes, the water diffusion and hydrogen bond structure is almost the same. This 
reflects the structure of the $(9,9)$ and $(16,0$) tubes illustrated in the Figure~\ref{fig_kneaded9} and ~\ref{fig_kneaded16}
respectively. The deformation changes the way water molecules connect and move inside the armchair nanotube from spiral-like to 
circular as the water melts inside the tube.

\section{Conclusions}

In this work, we analyzed the  diffusion coefficient of water under confinement in carbon nanotubes. Different tube sizes, 
topology and deformation were considered.

For the perfect nanotube the mobility of water in both 
armchair and zigzag topologies is very similar, with 
the exception of the $(9,9)$ and the $(16,0)$ cases. The water has a glassy state at $(9,9)$ while still moving at 
$(16,0)$, although both have the same diameter, which 
is attributed to the spiral and ring-shaped structure of the tubes respectively.

As the nanotube is deformed,  compression leads 
to a decrease in diffusion of water and an increase in the number of  hydrogen bonds between water molecules for both armchair 
and zigzag cases  for
the larger tubes, $(12,12)$ and $(21,0)$ respectively. For the smaller tubes,  $(9,9)$ and the $(16,0)$, the behavior is more 
complex since they form a single layer of water and the water-wall interactions become relevant. For the kneaded armchair $(9,9)$ 
nanotube, the deformation melts water, creating a structure similar to the observed in the $(16,0)$. 

Water molecules confined within deformed, wrinkled, and dented nanotubes show a decrease in mobility and an increase in the 
hydrogen bonding network. For the mobility water confined, the nanotube topology is relevant only for a specific diameter 
(about $1.22$ nm for the armchair and $1.24$ nm for the zigzag), for which for which the water confined at the perfect armchair 
$(9,9)$ nanotube features a glassy state. In this case, the deformation causes the water to come out of a state of almost 
zero mobility and leads to behavior similar to that observed in zigzag nanotubes.

\begin{acknowledgments}
This work is partially supported by Brazilian science agencies CNPq~--~through INCT-Fcx~--, CAPES, FAPEMIG, Universidade Federal 
de Ouro Preto, Universidade Federal do Rio Grande do Sul and Centro Nacional de Processamento de Alto Desempenho (CENAPAD).
\end{acknowledgments}

\bibliography{aipsamp}

\end{document}